\documentclass[journal=jacsat,manuscript=article]{achemso}
\usepackage{xcolor}

\usepackage[version=3]{mhchem} 

\author{Ajit Jena}
\author{Seung-Cheol Lee}
\email{seungcheol.lee@ikst.res.in}
\author{Satadeep Bhattacharjee}
\email{s.bhattacharjee@ikst.res.in}
\affiliation
{Indo-Korea Science and Technology Center, Bangalore, India}

\title[An \textsf{achemso} demo]
  {Surface-oxygen-passivation driven large anomalous Hall conductivity (AHC) in nitride MXenes: Can AHC be a tool to determine functional groups in 2D ferro(i)magnets?}

\abbreviations{IR,NMR,UV}

\begin{document}

\begin{tocentry}

\center
\includegraphics[keepaspectratio, height=4.5cm]{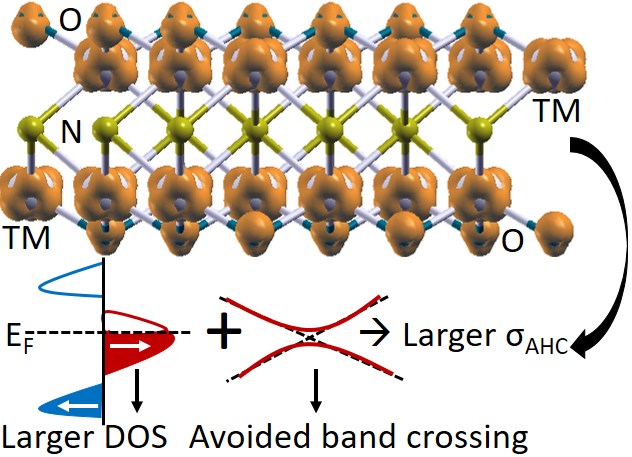}
In surface-oxygen-passivated nitride MXenes, relatively weak dispersive $O(p)-TM(d)$ hybridized states are found across $E_F$ and they seem to be degenerate at certain points in the BZ resulting larger uncompensated DOS. Hence, the combined effect of the larger DOS and avoided band crossings at $E_F$ induces larger AHC in the system.
\end{tocentry}

\begin{abstract}
 Identifying the existence of specific functional groups in MXenes is a difficult topic that has perplexed researchers for a long time. We show in this paper that in the case of magnetic MXenes, the magneto-transport properties of the material provide an easy solution. One of the fascinating properties that MXenes offer is the realization of intrinsic ferromagnetism which is important for two-dimensional (2D) materials family. The previous reports have only made a few statements on some MXenes citing its usefulness for spintronics related applications. Here, using first-principle calculations we have examined the actual magneto-transport phenomena in MXenes family.  We have considered all possible combinations of 3\textit{d} transition metals ($Ti, V, Cr$ and $Mn$) and nitride based functionalized $(O_2, F_2$ and $(OH)_2$) MXenes, $M_2NT_2$. The intrinsic anomalous Hall effect is investigated in $Cr$ and $Mn$ based MXenes as the compounds possess ground state stable ferromagnetic solutions. We demonstrate that intrinsic Anomalous Hall conductivity (AHC) can be used to identify the functional groups in MXenes.
 Additionally, half-metallic features of these ferromagnetic MXenes make them potential candidates for varieties of applications such as in logic and memory devices, quantum computations, spintronics etc. The maximum anomalous Hall conductivity (AHC) at Fermi energy, $E_F$, is found in case of $Mn_2NO_2$ (470 $S/cm$) which is attributed to the presence of avoided band crossing and larger density of states. Together, when considered all the studied systems, the AHC can be above 2500 $S/cm$ within $E_F \pm $ 0.25 $eV$. Our findings could be useful not only in guiding the experimentalists by considering AHC as a simple tool in determining the functional groups in 2D ferro(i)magnets, also, it could be useful in designing memory device with negligible stray fields.
\end{abstract}


The advances in the growth techniques in recent years provides plenty of scope to explore many possible functionalities in two-dimensional (2D) materials. Because of their ease of processing in promising device applications, they have attracted considerable attention in the scientific community. In this regard, graphene is at the forefront which has high mobile charge carriers having the dispersion curves similar to Dirac fermions with zero rest mass. In addition to graphene the other 2D materials which also receive interest for new-generation electronic and spintronic applications include hexagonal boron nitride and metal chalcogenides. However, after graphene, MXenes are the centre of attention as many of the other 2D systems remain as the subjects of academic interest only.

MXenes have the general formula of $M_{n+1}X_nT_x$, where $'M'$ is a early transition metal, $'X'$ can either be carbon or nitrogen or both and $'n'$ can vary from $1-4$ \cite{mxn1, mxn2, mxn-sci-rev}. In the above formula, $'T'$ stands for the functional groups such as $O, F$ and $OH$, their presence or absence respectively makes MXenes functionalized or bared. On the other hand, as far as the 3\textit{d} transition metal is concerned, combined experimental and theoretical studies are available up to $Mn$. There are reports on more than 30 different experimentally synthesized MXenes and more than a hundred bare Mxenes are predicted theoretically \cite{mxn-rise, mxn-sci-rev}. Experimentally, using hydrofluoric treatment, when 2D MXenes sheets are separated from its parent MAX phase one can not avoid the unintentional presence of functional groups in MXenes \cite{mxn1, mxn2, mxn-sci-rev}. Therefore, in order to have a more realistic scenario it is important to include the functional groups in the theoretical study. Unlike in experiments, it is possible to consider the effect of distinct functional groups in the theoretical calculations. Experimentalists denote the surface-terminating functional groups by a general symbol $T_x$ due to difficulties in detecting the presence of exact functional form. Therefore, since the exact terminating surface type is not clearly observed, it deems important to test with different surface-terminations in the theoretical study.

Magnetism is an important property especially for 2D systems because the spin degree of freedom of the electrons can easily be manipulated. There is always a challenge to achieve 2D magnetism intrinsically as most of the synthesized 2D materials are nonmagnetic. With the introduction of anisotropy or by doping incorporation or by quantum confinement, magnetism can be induced in nonmagnetic 2D systems \cite{induce-mag1, induce-mag2, induce-mag3, coatings}. However, such systems have limited practical applications as the enabled magnetism is generally not robust \cite{induce-mag3}. In this regard, few MXenes along with some monolayer magnets e.g., $CrI_3$ \cite{cri3} and $VSe_2$ \cite{vse2}, are becoming the research focus in the material science community due to their inherent magnetic property. 

Coming to MXenes, magnetism is mostly realized in bare surface cases due to unpaired $d$-electrons \cite{khazaei, mxn-25th-aniv}. With the presence of surface groups, the $M(d)$-orbitals form covalent bonding with the $T(p)$-orbitals and as a result the magnetism disappears (e.g. in $Ti_2CO_2$ , $Ti_2CF_2$ and $Ti_2C(OH)_2)$). However, there are exceptions: $Cr$-based carbides and nitrides are predicted to show magnetism with terminated surface groups ($T = O$, $OH$, or $F$) up to nearly room temperature \cite{khazaei, mxn-25th-aniv}. The present work is mainly on examining the intrinsic anomalous Hall effect (AHE) that occurs in ferro(i)magnetic systems driven by spin-orbit interaction. In this regard, the anomalous and ordinary Hall contributions are known to be proportional to the magnetization (\textbf{M}) and applied magnetic field (\textbf{H}) respectively \cite{kl1954, nagaosa}. This is in fact another reason to explore anomalous Hall effect in nitride MXenes rather than carbides as $'N'$ has one extra electron compared to $'C'$ \cite{kumar2017}. In any case, to the best of our knowledge, no such experimental or theoretical efforts have ever been made earlier in these intrinsic 2D magnets with a focus on AHE.


We have solved the Kohn-Sham Hamiltonian using pseudo-potentials (pp) and planewave basis sets as implemented in Quantum ESPRESSO (QE) density-functional theory (DFT) package \cite{giannozzi}. The exchange-correlation potential of the above Hamiltonian is approximated by PBE-GGA functional \cite{pbe} through optimized norm-conserving Vanderbilt pp \cite{hamann}. The kinetic energy cutoff for the planewave is taken as 80 $Ry$. Gaussian smearing of 0.005 $Ry$ is used both for the self-consistent (SC) and non-self-consistent (NSC) calculations. A tight energy threshold of $10^{-8} Ry$ has been set for the SC total energy calculations . The functionalized MXene sheets are sufficiently isolated from each other by vacuum ($c$ lattice constant = 20 \AA{}) in order to allow negligible interaction between the periodic units. The atomic positions and the in-plane lattice of all the structures have been fully relaxed. The magnetic ground states of every combination of  3\textit{d} transition metals and the functional groups are predicted from the collinear spin-resolved calculations. Further, the dynamical stability of all the AHE-investigated (ferromagnetic) systems is examined through density functional perturbation theory (DFPT) calculations.

We have used an efficient first-principles approach in which the Bloch wave functions are projected into maximally localized Wannier functions. Wannier90 tool (implemented within QE) has been used to compute the Wannier interpolated bands, Berry curvature and thereby the anomalous Hall conductivity (AHC) \cite{giannozzi, marzari, souza, pizzi}. The intrinsic AHC is proportional to the Brillouin zone (BZ) summation of the Berry curvature over all occupied states \cite{pizzi, kubler}

\begin{equation}
\sigma_{xy} (AHC) = \frac{e^2}{\hbar} \frac{1}{NV_c}\sum_{\textbf{k}\in(BZ)} (-1) \Omega_{xy}(\textbf{k}) f(\textbf{k}),
\label{eq:ahc}
\end{equation}

where the indices $x$ and $y$ are the Cartesian coordinates. $f(\textbf{k})$ stands for the Fermi distribution function, $\Omega_{xy}(\textbf{k})$ denotes the Berry curvature for the wave vector $\textbf{k}$, $N$ is the number of electrons in the crystal and $V_c$ is the unit cell volume. Very recently, using the same level of approach, we have reproduced the AHC values in different bulk Heusler compounds \cite{cts-prm, cfa-prb, mca-comm}. 

Spin-orbit coupling (SOC) is introduced in all the Berry curvature related calculations for which the direction of the magnetization has been set along the normal to the 2D surfaces. We find that the use of valence $d$ and $p$ orbitals as the projections in the Wannier90 calculations provides us very good interpolation. The Monkhorst-Pack \textbf{k}-grid of $16\times16\times1$ are considered in all the calculations, \textit{viz}.: SC, NSC and Wannier90. A denser \textbf{k}-grid of $150\times150\times1$ is taken to calculate the intrinsic AHC. The AHC value is found to be converged with the above \textbf{k}-grid choice. Through the  adaptive refinement scheme a fine mesh of 5 units are added around the points wherever the mode of the Berry curvature ($\mid{\Omega(\textbf{k})}\mid$) exceeds 100 bohr$^2$. The \textbf{q}-grid of $4\times4\times1$ has been taken to carry out the DFPT calculations.

\begin{table}[h!]
\centering
\caption{Total energy ($eV$) comparison of three spin configurations (FM, AFM and NM) for all possible combinations of different transition metals and functional groups. The energies are calculated with respect to the FM configuration of each functional group. The bold numbers represent the minimum energy among the three spin configurations.}
\begin{tabular}{|c c c c c c c c|} 
 \hline
& & & FM & & AFM & & NM  \\ [0.5ex] 
 \hline\hline
 $Ti\rightarrow$ & $O_2$  & & \textbf{0.0} & & 0.0004 & & 0.0004 \\ 
     & $(OH)_2$ & & 0.0 & & 0.0 & & \textbf{-0.0001} \\
     & $F_2$ & & 0.0 & & 0.0 & & \textbf{-0.00001} \\
 \hline
 $V\rightarrow$ & $O_2$  & & 0.0 & & \textbf{-0.112} & & 0.169 \\ 
     & $(OH)_2$ & & 0.0 & & \textbf{-0.166} & & 0.0002 \\
     & $F_2$ & & 0.0 & & \textbf{-0.079} & & 0.032 \\
 \hline
 $Cr\rightarrow$ & $O_2$  & & \textbf{0.0} & & 0.028 & & 1.7 \\ 
     & $(OH)_2$ & & 0.0 & & \textbf{-0.03} & & 1.83 \\
     & $F_2$ & & \textbf{0.0} & & 0.108 & & 2.41 \\
 \hline
 $Mn\rightarrow$ & $O_2$  & & \textbf{0.0} & & 0.25 & & 2.0 \\ 
     & $(OH)_2$ & & \textbf{0.0} & & 0.38 & & 2.71 \\
     & $F_2$ & & \textbf{0.0} & & 0.34 & & 3.5 \\
 \hline
 $Fe\rightarrow$ & $O_2$  & & \textbf{0.0} & & 0.223 & & 0.789 \\ 
     & $(OH)_2$ & & \textbf{0.0} & & 0.205 & & 1.91 \\
     & $F_2$ & & \textbf{0.0} & & 0.098 & & 2.21 \\
 \hline
 $Co\rightarrow$ & $O_2$  & & \textbf{0.0} & & 0.007 & & 0.014 \\ 
     & $(OH)_2$ & & \textbf{0.0} & & 0.174 & & 0.174 \\
     & $F_2$ & & \textbf{0.0} & & \textbf{0.0} & & 0.074 \\
 \hline
\end{tabular}
\label{table:ener}
\end{table}

\begin{table}[h!]
\centering
\caption{Obtained magnetic moments ($\mu_B$) in ferromagnetic MXnes which are considered for the AHE study in the present work.}
\begin{tabular}{|c c c c c c c|} 
 \hline
& & $O_2$ & & $(OH)_2$ & & $F_2$  \\ [0.5ex] 
 \hline\hline
 $Cr$ & & 5.0 & & 0.0 (AFM) & & 6.7 \\ 
 \hline
 $Mn$ & & 7.0 & & 7.35 & & 8.08 \\  
 \hline
\end{tabular}
\label{table:mm}
\end{table}

\begin{figure*}
\center
\includegraphics[scale = 0.5]{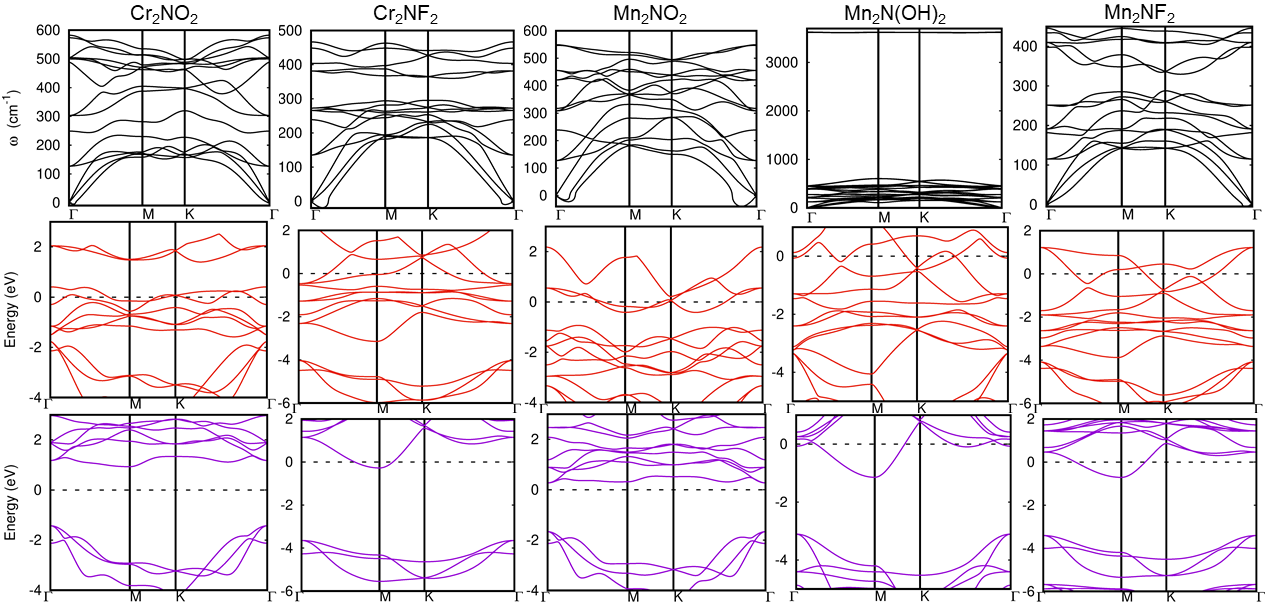}
\caption{Upper panel: Phonon dispersion relations of predicted five ferromagnetic MXenes considered for the AHE study of this work. Middle and lower panel respectively represents the band structure of spin-majority and spin-minority electrons. The Fermi energy ($E_F$) is set to 0 $eV$. Presence of large magnetic gap of $>$ 2 $eV$ (around $E_F$) in the spin-minority channel of each case suggests their half-metallic characteristics.}  
\label{fig:phonon-sp-bands}
\end{figure*}

At first, the first-principles total energy calculations are carried out considering the 3\textit{d} transition metals as $Ti, V, Cr, Mn, Fe$ and $Co$. For each transition metal, three possible surface terminations such as $O, OH$ and $F$ are chosen. The magnetic ground states have been predicted by comparing the total energies (per formula unit) of three different spin configurations, namely parallel (FM), anti-parallel (AFM) and no-spin (NM). Table.~\ref{table:ener} suggests that the $Ti$-based functionalized MXenes are found to be non-magnetic which is in agreement with the previous reports as discussed in the introduction. Though FM phase of $Ti_2NO_2$ possesses minimum energy the difference is extremely small ($<$ 1 $meV$) and the same is believed to occur due to certain numerical approximations. Therefore, like other two surface group cases ($(OH)_2$ and $F_2$) of $Ti$, the system can easily pair its spins and switch to NM configuration even with mild change in experimental conditions which are absent in our theoretical approach. The net magnetic moment of $Ti_2NO_2$ unit cell is also found to be very small (0.28 $\mu_B$). While all $V$-based nitride MXenes are predicted to show AFM as the ground state solution, all $Cr$ and $Mn$-based compounds are ferromagnetic except the case of $Cr_2N(OH)_2$. We have also included the case studies of $Fe$ and $Co$ which too show FM ground state with any surface group. It should be noted that there is minimal energy difference between three magnetic configurations of $Co_2NO_2$ and in case of $Co_2NF_2$, the FM and AFM orderings have equal energy. As we are interested in robust ferromagnetic systems (with majority of surface groups) we have picked up $Cr$ and $Mn$ based MXenes to examine the anomalous Hall effect. The calculated magnetic moments (per formula unit) of these ferromagnets are provided in Table.~\ref{table:mm}. Another reason of not considering MXenes beyond $Mn$ for further investigation is due to lack of experimental realization. The ferromagnetic ground states in $Cr_2NO_2$ and $Mn_2NT_2$ with any surface group have also been reported earlier where more discussions on the ground state magnetism are presented \cite{wang2016, kumar2017}.

Table.~\ref{table:ener} further suggests that MXenes with four or more (five) number of valence \textit{d}-electrons mostly possess FM as the ground state. We believe that this observation is one of its first kind made in MXenes which could be useful in designing stable intrinsic 2D ferromagnets. $Cr$ and $Mn$ respectively have four and five atomic valence \textit{d}-electrons. $Fe$ is equivalent to the case of $Cr$ case with $5\uparrow-1\downarrow = 4$ unpaired \textit{d}-electrons. Similarly, $V$ and $Co$ are equivalent to each other having 3 unpaired \textit{d}-electrons in which ground state AFM solution is mostly favourable. In the competition between magnetism and $p-d$ covalent interaction, long range ferromagnetic ordering is expected to be predominant in MXenes with more number of unpaired \textit{d}-electrons ($Cr$ and $Mn$ based compounds). The dynamical stabilities of five ferromagnetic MXnenes have been examined by calculating the phonon dispersion relations (upper panel of Fig.~\ref{fig:phonon-sp-bands}). The absence of imaginary frequencies in $Cr_2NO_2$, $Mn_2N(OH)_2$ and $Mn_2NF_2$,  and in case of $Cr_2NF_2$ and $Mn_2NO_2$, its existence (minimal) only near the zone center suggest that these ferromagnetic MXenes are dynamically stable. It would be interesting to explore detailed phonon assisted properties such as phonon-electron coupling etc. in these compounds. It is worth mentioning that all the five FM MXenes are also found to be half-metallic (please see the band structures in Fig.~\ref{fig:phonon-sp-bands}). As it can be seen there exist a substantial magnetic gap of at least 2 $eV$ (near $E_F$) in the spin-minority channel of each case. Half-metallic ferromagnets are useful for spintronic applications in which the carrier transport will be contributed by only one spin-channel (majority in the present case) making both electrons and holes as 100 $\%$ spin-polarized.

The origin of anomalous Hall effect in ferro(i)magnetic materials is explained by two possible mechanisms namely, extrinsic and intrinsic \cite{nagaosa}. The extrinsic mechanism further includes the contributions from skew scattering and side jump mechanism and is related to the scattering events. However, the intrinsic AHE is correlated to the electronic band structure of the material. Karplus and Luttinger \cite{kl1954}, long ago in a classic work, have shown that the applied electric field combines with spin-orbit coupling to give an additional term (which is related to the Berry curvature) added to the usual electron velocity. This additional term is known to be anomalous velocity.
Though the intrinsic mechanism is purely of the band structure origin, due to certain scaling law, practically it is not possible to decouple the intrinsic and side jump mechanism and the latter has very small contribution to the AHC in comparison to the intrinsic part \cite{cfa-prb}. The present work discusses the intrinsic anomalous Hall conductivity which has contributions only from Berry curvature which can be expressed  (for the n$^{th}$ band) as follows~\cite{Allen},
\begin{equation}
\Omega^n_{xy}=\sum_{m\neq m}~\frac{Im[<\psi_{nk}|\hat v_x|\psi_{mk}><\psi_{mk}|\hat v_y|\psi_{nk}>]}{(\varepsilon_{nk}-\varepsilon_{mk})^2}
\label{BC}
\end{equation}
The Berry curvature was calculated from the tight-binding model Hamiltonian constructed by projecting the Bloch wave functions to Wannier functions. $\varepsilon_{nk}$ is the eigenvalue of the n$^{th}$ eigen state, $|\psi_{nk}>$ . The velocity operators $\hat v_{x(y)}$ are constructed from the tight-binding Hamiltonian, $\hat H$, via $\hat v_{x(y)}=\frac{1}{\hbar}\frac{\partial \hat H}{\partial k_{x(y)}}$. The large contribution to the Berry curvature and hence the AHC comes from the situations when the eigenvalues $\varepsilon_{nk}$ and $\varepsilon_{mk}$ are very close, particularly in the case of avoided band crossings.

\begin{figure*}
\center
\includegraphics[scale = 0.9]{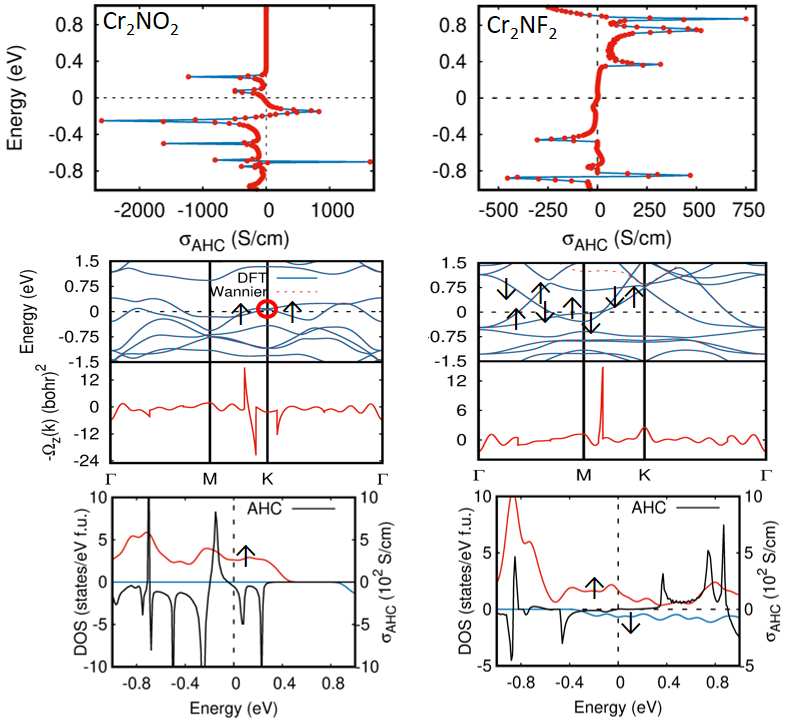}
\caption{\textbf{Top}: Variation of anomalous Hall conductivity (AHC) as a function of energy for $Cr_2NO_2$ and $Cr_2NF_2$. The $E_F$ is set to 0 $eV$. \textbf{Middle}: Full electronic band structures (DFT, blue solid lines) of corresponding compounds calculated with the introduction of spin-orbit coupling. The Wannier interpolated bands (red dotted lines) are also shown in the same plot. Since we have achieved very good interpolation one curve is exactly sitting on the other, especially below $E_F$. The red circle at $E_F$ marks the avoided band crossing (occurs due to same spin) which gives rise to non-zero Berry curvature in the same region of the $k$-space. The Berry curvature calculated along the same high-symmetry path is also shown below of the band structure. \textbf{Bottom:} The AHC curves are shown again along with the spin-$\uparrow$ (red curve),$\downarrow$ (blue curve) density of states. It can be understood that in case of $Cr_2NF_2$,  due to compensation effect of spin-$\uparrow$,$\downarrow$ contributions, the net AHC is close to zero. }
\label{fig:cr-ahc}
\end{figure*}

The intrinsic AHC associated results of $Cr$-based compounds are shown in  Fig.~\ref{fig:cr-ahc}. As discussed earlier, $Cr_2N(OH)_2$ is found to be AFM whereas $Cr_2NO_2$ and $Cr_2NF_2$ show ferromagnetic behaviour and therefore, the anomalous Hall transport studies have been performed on the latter two cases. 

The Berry curvatures plotted along the high-symmetry paths of hexagonal lattice are shown at the bottom of Fig.~\ref{fig:cr-ahc}. The curvature is calculated from the maximally localized Wannier functions. These Wannier orbitals are also compared with the spin-orbit coupled band structures. The figure suggests that excellent matching between the two have been obtained. Therefore, with this same set of parameters, one would predict accurate properties that can be obtained from Wannier functions. One of such quantities is the intrinsic AHC which is shown as a function of $E_F$ at the top panel. The conductivity values are 2D-corrected using the relation,

\begin{equation}
\sigma_{xy}^{2D} (AHC) = \sigma_{xy} (AHC) \times \frac{c_{lat}}{c_{t}^{real}},
\label{eq:ahc-2d}
\end{equation}

where $\sigma_{xy} (AHC)$ is the AHC value obtained from our first-principles calculation, $c_{lat}$ is the $c$ lattice constant (= 20 \AA{}) including the vacuum and $c_{t}^{real}$ is the actual thickness of the functionalized MXene sheet. Van der Waals radii of the terminated atoms are taken into account while calculating the thickness of the MXene sheets. The above correction takes care of the accurate unit cell volume ($\Omega_c$) that exists in the denominator of Eq.~\ref{eq:ahc}.    

The magnitude of AHC in $Cr_2NO_2$ is found to be 51 $S/cm$ and is negligible (nearly zero) in case of $Cr_2NF_2$. The non-zero AHC across $E_F$ in the former case is attributed to the Berry curvature spikes that occur in presence of avoided band crossings or near-degeneracies \cite{bc-ahc-org}. The avoided band crossing at $E_F$ is marked by the red circle. No such crossings or degeneracies, between two bands having same sign of spin, are found at $E_F$ in $Cr_2NF_2$. The bottom panel of Fig.~\ref{fig:cr-ahc} suggests that the AHC in $Cr_2NO_2$ is exclusively contributed by the spin-majority electrons. Whereas, in $Cr_2NF_2$, the negligible AHC is the outcome of the compensation effect involving spin-majority and spin-minority electrons. Whether the contributions are from the same or different spin types, that can be understood while looking at the Fermi-bands demonstrated by the up and down arrows. The spin-signs of the Fermi-states can also be identified with the help of spin-polarized band structures shown in Fig.~\ref{fig:phonon-sp-bands}. Other avoided crossings away from the Fermi energy contributes to the appearance of additional Berry curvature spikes along the high symmetry path. Further, the AHC in case of $Cr_2NO_2$ abruptly increases up to 2600 $S/cm$ at $-$ 0.25 $eV$ which can be realized by electric gating or by hole doping.   

\begin{figure*}
\center
\includegraphics[scale = 0.6]{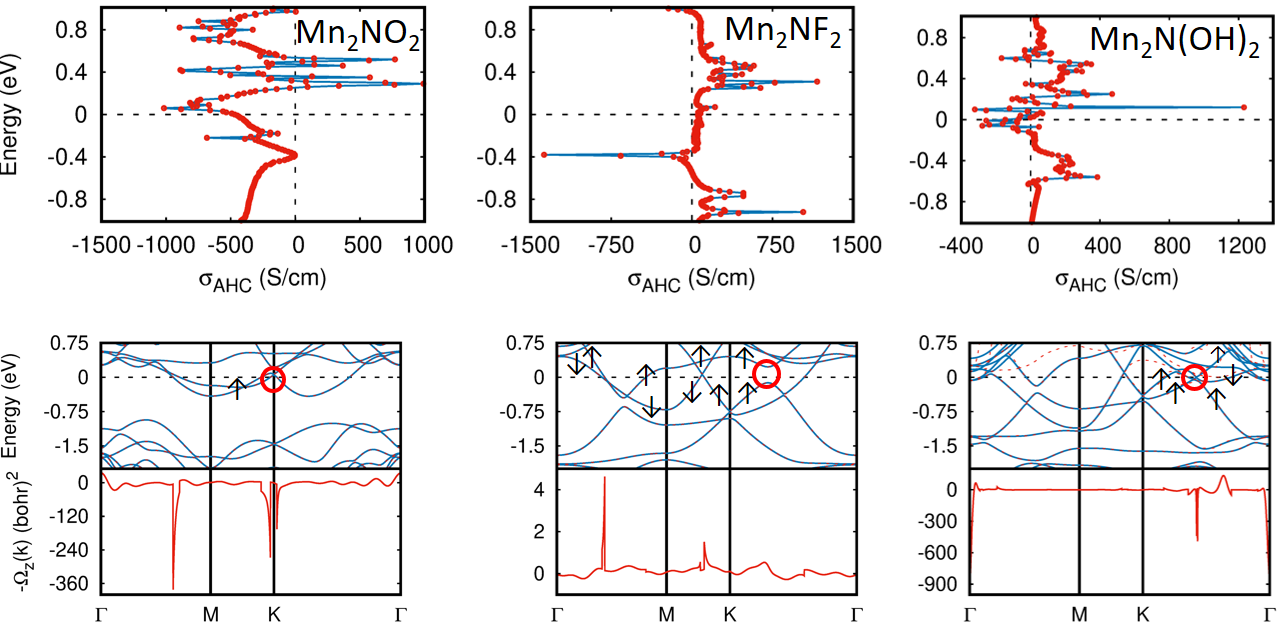}
\caption{Similar plots (as shown in top two panels of Fig.~\ref{fig:cr-ahc}) are presented for $Mn$-based MXenes.}
\label{fig:mn-ahc}
\end{figure*}

In Fig.~\ref{fig:mn-ahc}, similar results are presented for $Mn$-based MXene which all possess ferromagnetic ground state with any surface group. The Fermi-energy AHC for $Mn_2NO_2$, $Mn_2NF_2$ and $Mn_2N(OH)_2$ are found to be 470 $S/cm$, 66 $S/cm$ and 147 $S/cm$ respectively. Like in the $Cr$ case, we find larger AHC in $Mn_2N$ when the surface is passivated with oxygen.

Now, we discuss about the variation of AHC as a function of energy which are presented at top panel of Fig.~\ref{fig:mn-ahc}. For an energy extremely close to $E_F$ ($\sim$ 0.05 $eV$), the AHC of $Mn_2NO_2$ reaches up to 1000 $S/cm$. Similarly, $Mn_2N(OH)_2$ provides AHC of above 1200 $S/cm$ at 0.1 $eV$. However, within $E_F$ $\pm$ 0.25 $eV$, the maximum AHC of $Mn_2NF_2$ is found to be 219 $S/cm$. As we saw, even for same transition metal, the AHC across $E_F$ can be significantly different with different surface terminations. Therefore, it can be hypothesized that the magnitude of AHC does not explicitly follow any strict rule as far as the transition metal is concerned. Rather, it depends on the electronic band structure of the MXenes with different functionalizations which can have dissimilar avoided band crossings or near-degeneracies at or away from $E_F$. One thing is common in both the transition metal cases is that the surface-oxygen-passivated MXene induces larger AHC whereas fluorine functionalization provides least value. 

\begin{figure*}
\center
\includegraphics[scale = 0.6]{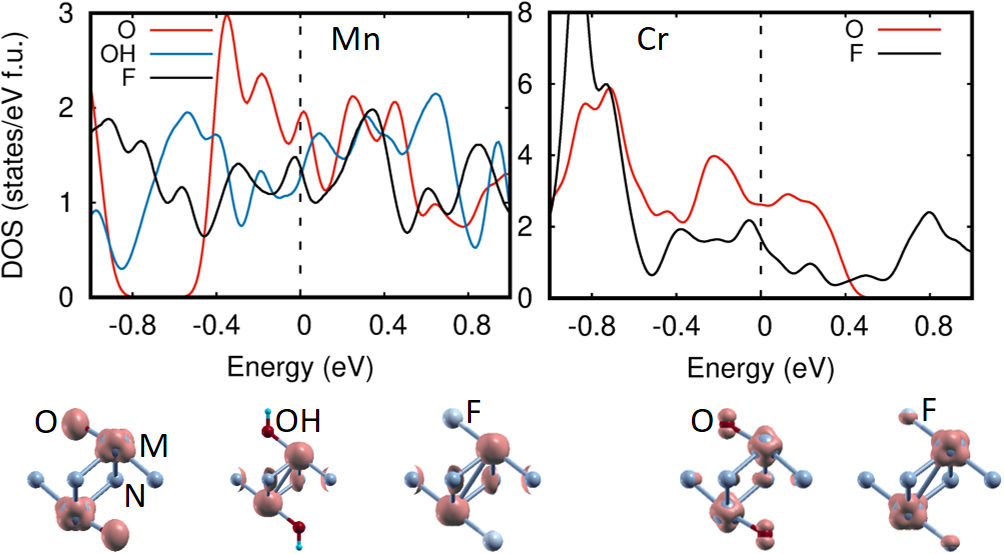}
\caption{Spin-majority density of states and charge density plots of ferromagnetic MXenes with different surface functionalizations.}
\label{fig:dos-cd}
\end{figure*}

As discussed earlier, the avoided band crossings or near-degeneracies at certain energy contribute to the non-zero AHC and the AHC in connection to the avoided crossings is already discussed through Fig.~\ref{fig:cr-ahc} and Fig.~\ref{fig:mn-ahc}. Now, to substantiate our understandings of the effect of near-degeneracies on the magnitudes of AHC for different surface groups we plot the density states (DOS) in Fig.~\ref{fig:dos-cd}. Since the AHC in the present cases is mostly contributed by the spin-majority electrons the DOS of the same have been presented. We can see from Figs.~\ref{fig:phonon-sp-bands} to ~\ref{fig:mn-ahc} that, the $E_F$ is occupied by relatively weak dispersive states when the MXene surface is terminated by oxygens. The energy of those relatively localized electrons seems to be degenerate at some points close to $E_F$. Certainly, the DOS at $E_F$ in corresponding MXenes will be larger compared to others and the Berry curvature will stand uncompensated due to half-metallic feature. Interestingly, the AHC trend for different functionalizations in both the transition metal cases follows the DOS at $E_F$. Though $Mn_2N(OH)_2$ and $Mn_2NF_2$ seem to meet at same point the integrated DOS in the vicinity of $E_F$ is larger in the former case and hence, it provides larger AHC both at $E_F$ and close-by. Therefore, the combined effect of larger magnetization, larger DOS and the presence of avoided band crossing at $E_F$ drives $Mn_2NO_2$ to generate larger AHC in the system.

\begin{figure*}
\center
\includegraphics[scale = 0.6]{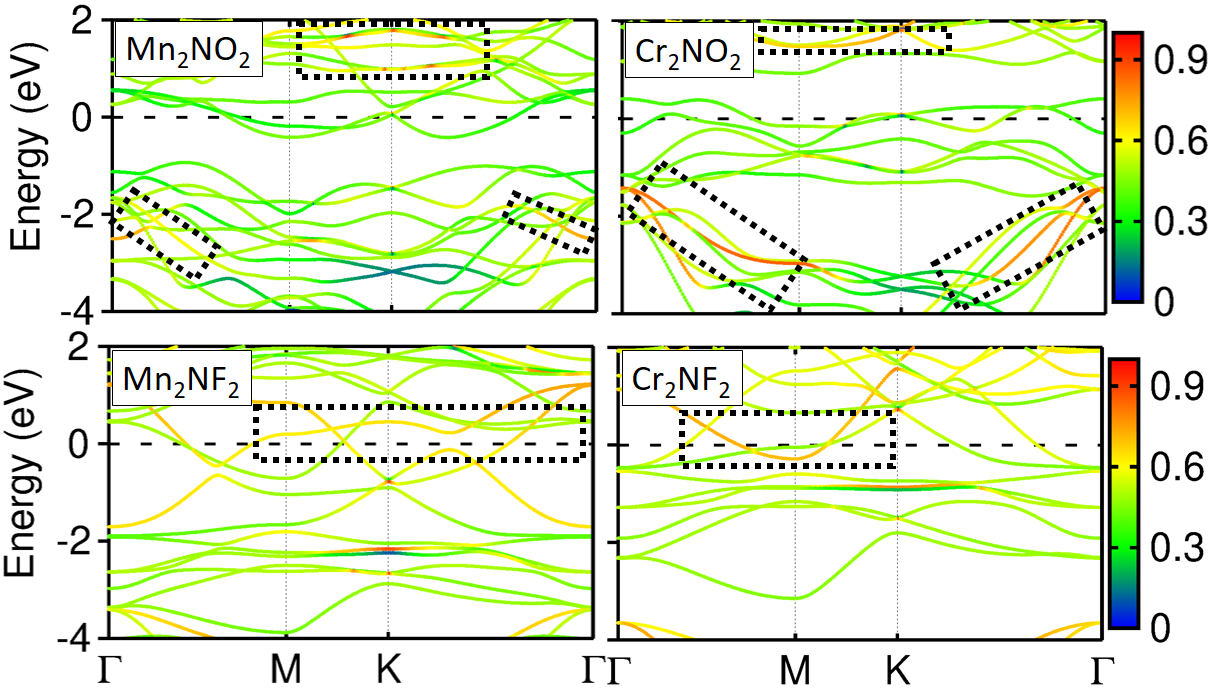}
\caption{Comparison of projected $N(p)-M(d)$ hybridized states in \textit{Cr} and \textit{Mn} based nitride MXenes for \textit{O} and \textit{F} surface groups. Dotted rectangles mark the portion of the band structure where the $N(p)-M(d)$ interaction is substantial ($\sim$ 0.8 in the colorbox). The same occurs at $E_F$ in case of fluorine surface termination and in the case of oxygen as surface group, those hybridized states are found away from the Fermi energy.}
\label{fig:proj-bands}
\end{figure*}

The charge-density visualizations shown at the bottom of Fig.~\ref{fig:dos-cd} demonstrate the atoms that contribute to the states at $E_F$. Not surprisingly, the transition metals show greater affinity towards oxygen and the metal-oxygen ($M-O$ = 1.9 \AA{}) bond length (only in oxygen-passivated cases) is $\sim 10 \%$ lesser than that of metal-nitrogen ($M-N$) and metal-fluorine ($M-F$) bond lengths. On the other hand, due to $O-H$ bonding in $Mn_2N(OH)_2$, the $M-O$ bond length is increased and equals ($\sim$ 2.1 \AA{}) to that of $M-N$ and $M-F$. Usually, the states that arise due to $M-O$ coordinated covalent interaction are more dispersed as happens in many transtion metal oxides \cite{imada}. However, the disorders in the system (in the form of structural/chemical) would make the states localized \cite{imada, lmpo}. The localization in the present case occurs due to dimensional confinement imposed on the $p$-electrons of terminating oxygens. Nevertheless, whenever the $N(p)-M(d)$ covalent interaction is predominant, the geometrical position of $N$ allows the $N(p)$ electrons to move freely in all three directions and the states are found to be relatively dispersive. The $N(p)-M(d)$ hybridized states at $E_F$ are formed in the absence of oxygen surface group. In oxygen-passivated systems, the $N(p)-M(d)$ hybridized states are found deep in the valence and conduction band energy (see Fig.~\ref{fig:proj-bands}) and the weakly dispersive $O(p)-M(d)$ hybridized states at $E_F$ hold the accountability for larger AHC in $M_2NO_2$.   

The AHC $vs$ $E_F$ plots further suggest that the conductivity in the stoichiometric compounds can be tuned substantially with electron/hole doping which can be achieved by means of foreign element doping. Even without incorporating foreign elements, intrinsically it is easier to shift $E_F$ in 2D system by applying gate voltage since the carriers are freely available on the surfaces. The predicted large intrinsic AHC, especially in surface-oxygen-passivated MXenes, suggests their usefulness for memory device applications in which the detrimental effect of the perturbing fields can obviously be minimized.

Our findings make another important remark for the ferromagnetic MXene family in particular. As discussed in the introduction it is difficult to determine the presence of exact functional group with the help of existing experimental techniques. Lack of know-how about the actual composition hinders our control to tune the materials properties in an efficient way. Since, experimentally, it is not difficult to measure the AHC one can map the experimental value with the theoretically computed ones to realize the actual functionalized form (individual or combination of different surface groups). Therefore, in this regard, the theoretical results presented here are concluded with respect to their implications for experimental investigation.

In summary, carrying out Ab \textit{initio} calculations, we show that the 2D intrinsic half-metallic and ferromagnetic MXenes possess wide range of intrinsic AHC ($0-2600$ $S/cm$) within $E_F \pm $ 0.25 $eV$. There is no general rule which can explain AHC in connection with the involved transition metal, rather it purely depends on the electronic band structure of the involved surface group. The anomalous Hall transport results of the studied half-metals could be useful to engineer efficient planar storage devices. Further, we conclude with an inquisitiveness to envisage whether AHC can be a tool to detect the functional groups in ferro(i)magnets.

\begin{acknowledgement}

A. Jena acknowledges the computational and financial support from IKST.

\end{acknowledgement}




\bibliography{MXenes-AHC-ACS-V3}

\end{document}